\documentclass[1p]{elsarticle}
\usepackage{graphicx,t1enc}
\usepackage{amssymb,amsmath}
\usepackage[utf8]{inputenc}
\usepackage{epsf}
\usepackage{natbib}

\begin{document}
\begin{frontmatter} 
\title{Oscillations from satiation of predators}

\author[conicet,cab,ib]{M. N. Kuperman}
\ead{kuperman@cab.cnea.gov.ar}
\author[conicet,cab]{M. F. Laguna}
\ead{lagunaf@cab.cnea.gov.ar}
\author[conicet,cab,ib]{G. Abramson}
\ead{abramson@cab.cnea.gov.ar}
\author[conicet,fb]{A. Monjeau}
\ead{amonjeau@fundacionbariloche.org.ar}
\author[iidypca]{J. L. Lanata}
\ead{jllanata@conicet.gov.ar}

\address[conicet]{Consejo Nacional de Investigaciones Cient\'{\i}ficas y T\'ecnicas}
\address[cab]{Centro At\'omico Bariloche (CNEA), R8402AGP Bariloche,  
Argentina}
\address[ib]{Instituto Balseiro, Universidad Nacional de Cuyo, Argentina}
\address[fb]{Fundación Bariloche, R8402AGP Bariloche, Argentina}
\address[iidypca]{Instituto de Investigaciones en Diversidad Cultural y Procesos de Cambio, CONICET-UNRN, R8400AHL Bariloche, Argentina}

\begin{abstract}
We develop a mathematical model of extinction and coexistence in a generic predator-prey ecosystem 
composed of two herbivores in asymmetrical competition and a hunter exerting a predatory pressure on 
both species. With the aim of representing the satiety of hunters when preys are overabundant, we 
introduce for the predation behavior a dependence on preys density. Specifically, predation is 
modeled as growing proportionally to the presence of herbivores at low density, and saturating when the total population of prey is sufficiently large. The model predicts the 
existence of different regimes depending on the parameters considered: survival of a single species, 
coexistence of two species and extinction of the third one, and coexistence of the three species. 
But more interestingly, in some regions parameters space the solutions oscillate in time, both as a transient phenomena and as persistent oscillations of 
constant amplitude. The phenomenon is not present for the more idealized linear predation model, suggesting that it can be the source of real ecosystems oscillations. 
\end{abstract}

\date{\today}


\end{frontmatter}

\section{Introduction}

The use of mathematical models in biology in general and in ecology in particular has grown 
significantly in the last decade. This is due in part to their predictive capacity, but also due 
to their power to order and systematize assumptions and thus contribute to 
elucidate the behavior of complex biological systems. In fact, the interrelation of factors as 
diverse as climate, access to resources, predators and human activity, makes it necessary to 
develop mathematical models that allow predicting the effect of each of them on the species 
involved, showing possible scenarios of coexistence or extinction. A large number of 
publications on topics such as predator-prey models \cite{1,2,3}, intra- and inter-specific 
competition \cite{4,5,6}, or  habitat fragmentation \cite{7,8,9} can be found, but more research is 
still needed on how to integrate all these mechanisms together.

In previous works  we developed a mathematical model of extinction and coexistence in a generic 
predator (or hunter)-prey ecosystem. In order to characterize the general behaviors we focused on a trophic 
network of three species: two herbivores in asymmetric competition and one predator \cite{10,11}. 
This problem was studied by means of ordinary differential equations and stochastic simulations. 
Both approaches provided similar and interesting results. The model predicts the existence of 
different regimes depending on the parameters considered: survival of one species, coexistence of 
two and extinction of the third (in the three possible combinations), and coexistence of the three 
species involved \cite{10}. Moreover, the results presented in \cite{11} indicate that the superior 
competitor of the hierarchy is driven to extinction after the introduction of hunters in the model. This 
happens even in pristine habitats (with no environmental degradation) and, more relevantly, even if 
the predatory pressure is higher on the inferior herbivore. 

In the original model we proposed that predation grew proportionally to the density of herbivores. 
While this approach is valid for ecosystems with low density of preys, it introduces an unrealistic 
behavior of the predator population when the density of available prey is 
high. The model implicitly assumes that the predator or hunter never quench, even when there is an 
overabundance of prey. In the present work, and in search of a better representation of predation, 
we analyze a variation of this model. Satiety of the predator or hunter is incorporated in the 
mathematical description as an asymptotic saturation in the term of predation.

The analysis of this model shows new results. The most interesting aspect of the solutions is the 
temporal oscillation of the populations. Under certain conditions these oscillations are transient 
and decay to a stable equilibrium. But in other situations oscillations are maintained 
indefinitely. In fact, we found regions of coexistence of the three species with persistent 
oscillations of constant amplitude. These dynamic regimes enrich the predictive properties of the 
model, so we expect our results to drive the search for evidence of oscillations in populations of 
current and extinct species.

In the Section~\ref{model} we introduce the mathematical model. Section~\ref{results} is devoted 
results, whereas in Section~\ref{conclusion} we discuss the main implications of the results 
and  possible future directions.

\section{Model with saturation in predation }
\label{model}

Our dynamical model requires a set of rules determining the temporal evolution of the system. These 
rules are inspired by the life history and the ecological interactions of the species involved, 
corresponding to biotic, environmental and anthropic factors~\cite{10}. In order to gain insight 
into the possible outcomes of different scenarios of interest, we have intentionally kept our system 
relatively simple: two herbivores in a hierarchical competition and a hunter exerting a predatory 
pressure on both. Some details of the ecological implications are discussed below. 

The original model, proposed in \cite{10}, can be described by the following set of equations: 
\begin{subequations}
\label{eq:or}
\begin{align}
\frac{dx_1}{dt}&=c_1 x_1(1-x_1)-e_1 x_1-\mu_1 x_1 y,  \label{eq:ora}\\
\frac{dx_2}{dt}&=c_2 x_2(1-x_1-x_2)-e_2 x_2-\mu_2x_2 y,  \label{eq:orb}\\
\frac{dy}{dt}&=c_y y(x_1+x_2-y)-e_y y. \label{eq:orc}
\end{align} 
\end{subequations}

Each species is described by a dynamical variable: $x_1$ and $x_2$ are the herbivores, and $y$ is 
the predator. Eqs.~(\ref{eq:or})  describe the time evolution of these variables. We imagine that 
both herbivores feed on the same resource and therefore compete with each other. This is represented 
by the first terms of Eqs.~(\ref{eq:ora}-\ref{eq:orb}). 

Such competition is asymmetrical, as it happens in most natural situations. This has interesting 
consequences, since coexistence under these circumstances requires advantages and disadvantages of 
one over the other. Consider, for example, that the individuals of each species are of different 
size, or temperament, such that species $x_1$ can colonize any available patch of 
habitat, and even displace $x_2$, while species $x_2$ can only occupy sites that are not already 
occupied by $x_1$. In this regard, we call $x_1$ the superior or dominant species of the hierarchy, 
and $x_2$ the inferior one. Since body size is often the main factor establishing this hierarchy, we 
imagine that $x_1$ is larger than $x_2$. This asymmetry is reflected in the logistic terms 
describing the competition in Eqs.~(\ref{eq:ora}) and (\ref{eq:orb}), as $x_1$ limits the growth of 
$x_2$ in Eq.~(\ref{eq:orb}), while the reciprocal in not true. This mechanism can be considered as a 
weak competitive displacement. A strong version of the competitive interaction between the 
herbivores was also considered in our previous works, consisting of an additional term $-c_1 x_1 
x_2$ in Eq.(\ref{eq:orb}). 

In other words, we have intra-specific competition in both species, but only $x_2$ suffers from the 
competition with the other species, $x_1$. In this context, for $x_2$ to survive requires that they 
have some advantage other than size, typically associated with a higher reproductive rate or a lower 
need of resources.

Besides, the equations for the herbivores also include a mortality term with coefficient $e_i$  and 
a predation term with coefficient $\mu_i$. The equation for the predator $y$ is also logistic, with 
a few differences. The reproduction rate of the predator is limited by intra-specific competition 
but enhanced by the presence of prey. 

Now we analyze a variation of the previous model, seeking a more realistic representation of the predation (or hunting) term. The differential equations in the model with saturation are as follows:
\begin{subequations}
\label{eq:mod}
\begin{align}
\frac{dx_1}{dt}&=c_1 x_1(1-x_1)-e_1 x_1-\frac{\mu_1 x_1y}{x_1+x_2+d_1},  \label{eq:moda}\\
\frac{dx_2}{dt}&=c_2 x_2(1-x_1-x_2)-e_2 x_2-\frac{\mu_2 x_2 y}{x_1+x_2+d_2},  \label{eq:modb}\\
\frac{dy}{dt}&=c_y y(x_1+x_2-y)-e_y y. \label{eq:modc}
\end{align} 
\end{subequations}

Observe that, while the predation terms undermine the population of herbivores, predation does not 
grow proportionally to the presence of prey, but rather saturates if the combined prey population is 
sufficiently large. This represents a satiation if preys are overabundant. Note that two new 
parameters $d_1$ and $d_2$ are included in Eqs.~(\ref{eq:mod}), representing the departure from 
proportionality.

In the next section we present the main results of the model described by Eqs.~(\ref{eq:mod}) and compare them with those of Eqs.~(\ref{eq:or}).

\section{Results}
\label{results}

\begin{figure}[t]
\centering
\includegraphics[width=0.7\columnwidth]{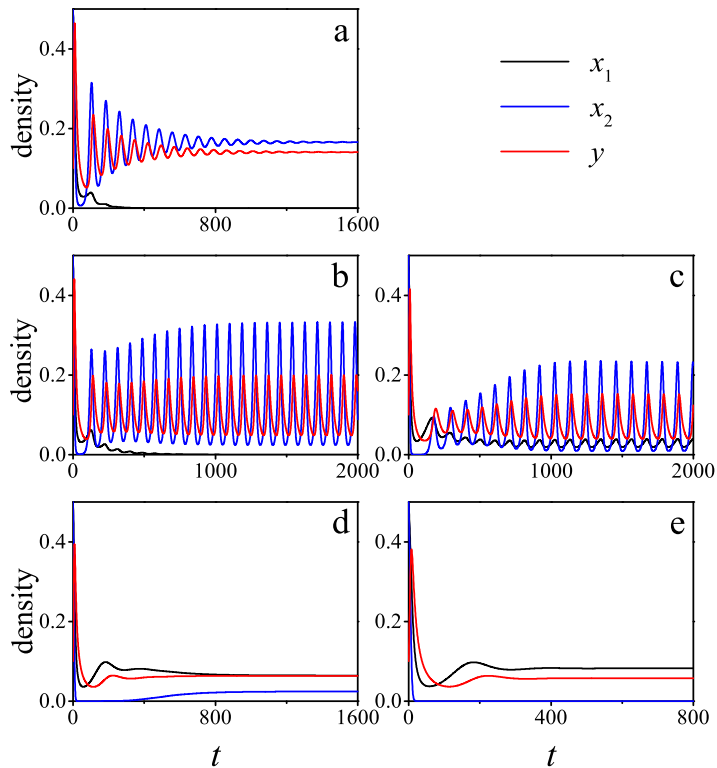}
\caption{Temporal evolution of the species populations for different predations pressures over 
$x_2$ (a) $\mu_2$=0.33, (b) $\mu_2$=0.40, (c) $\mu_2$=0.50, (d) $\mu_1$=0.60, (e) $\mu_2$=0.67. Other parameters remain fixed: $c_1$=0.14, $c_2$=0.2, $c_y$=0.4, $e_1$=0.2, $e_2$=0.015, $e_y$=0.01, corresponding to coexistence 
in the absence of predators pressure, and $d_1$=0.22, 
$d_2$=0.14,  $\mu_1$=0.14 < $\mu_2$, indicating a higher predation pressure on $x_2$.}
\label{evol}
\end{figure}

The study of the system with saturation of the predating pressure shows interesting and much richer 
results than those of our previous models. While the model described by Eqs.~(\ref{eq:or}) predicts 
several different regimes, with three and two  species coexistence, the steady state solutions are 
always stable nodes. Here we show that the saturation effect induces oscillatory solutions.

Without losing generality, we have restricted the values of the parameters within a range that 
shows all the behaviours displayed by the model, especially those scenarios of coexistence between 
two or all three species. The parameters are chosen in such a way that in the absence of predation 
pressure ($\mu_1$ = $\mu_2$ = 0), a coexistence of the three species is achieved. Moreover, the 
predation pressure over $x_1$ is kept fixed at a value $\mu_1$ < $\mu_2$, corresponding to 
situations where the inferior species is hunted more frequently than the superior one. 

We plot in Fig.~\ref{evol} the temporal evolution of the population densities for different values of $\mu_2$, 
the predation pressure over the inferior herbivore, $x_2$. The first panel (Fig.~\ref{evol}a) shows the behavior of the populations when a relatively 
low predation pressure is exerted on $x_2$, $\mu_2$=0.33. In this case we observe damped oscillations, which converge to the extinction of the superior herbivore $x_1$ and to the coexistence 
of the other two species, the predator $y$ and the inferior herbivore $x_2$. A higher value of  
$\mu_2$=0.40 is not enough to allow the survival of $x_1$ but produces sustained oscillations of 
$y$ and $x_2$ (see Fig.~\ref{evol}b). An even higher pressure on $x_2$ 
($\mu_2$=0.50) and the equilibrium between herbivores is achieved, and the three species coexist.  This is shown in Fig.~\ref{evol}c, with persistent oscillations of constant amplitude. If we increase further the predation pressure on $x_2$, the  oscillations disappear. Still, the coexistence of the three species is possible, as shown in  Fig.~\ref{evol}d. As expected, a larger predation pressure on the inferior herbivore will finally produce its  extinction, as seen in Fig.~\ref{evol}e.  As mentioned before, these non-oscillating behaviors were also observed in our previous model.

\begin{figure}[t]
\centering
\includegraphics[width=0.8\columnwidth]{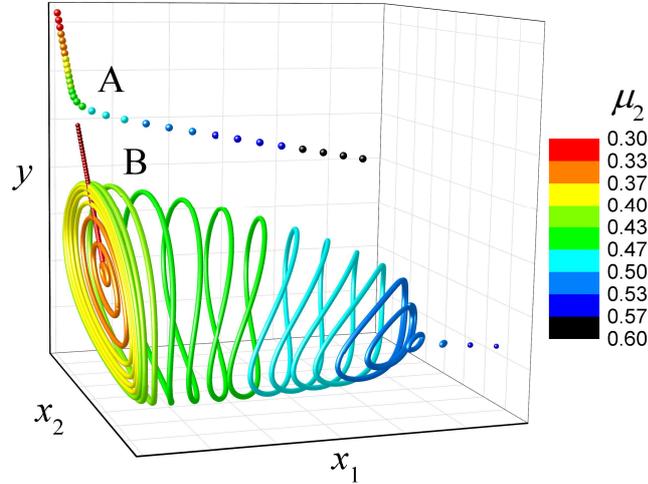}
\caption{Asymptotic solutions for a range of values of $\mu_2$, with A) corresponding to Eqs.~(\ref{eq:or}) (no satiety) and B) corresponding to Eqs.~(\ref{eq:mod}) (predation saturation). All the remaining parameters are equal to those 
of Fig.~\ref{evol}.}
\label{ciclos}
\end{figure}

In order to provide a visual representation of the steady state behavior of both  systems, 
Eqs.~(\ref{eq:or}) and (\ref{eq:mod}), we show in Fig.~\ref{ciclos} the stable equilibria and limit 
cycles corresponding for the solutions of both models, for a range of $\mu_2$ and the same choice of 
the values of the rest of the parameters as in Fig.~\ref{evol}.
On the one hand the asymptotic solutions corresponding to the model described by 
Eqs.~(\ref{eq:or}), without saturation in the predation, converge to stable nodes, showing three 
species coexistence for all the values of $\mu_2$ displayed. These are the set of solutions 
indicated as A on Fig.~\ref{ciclos}. 

On the other hand, the steady state solutions of Eqs.~(\ref{eq:mod}) show both stable nodes and cycles. These are indicated as B on Fig.~\ref{ciclos}. The dynamics of the cycles is rather interesting. 
For $\mu_2\lesssim 0.36$ we have non-oscillatory solutions, nodes located on the vertical $(x_2,y)$ 
plane, that appear as an oblique line of dots in Fig.~\ref{ciclos} on the left of the plot. In this 
regime the dominant herbivore, despite of being less predated on than the inferior one, can not 
persist. 
At $\mu_2\approx 0.36$ there is a Hopf bifurcation and cycles (still on the vertical $(x_2,y)$ 
plane) appear. 
Then, at $\mu_2 \approx 0.46$ a new bifucartion occurs. This time it is a transcritical bifurcation 
of cycles, as will be shown later. The superior herbivore can now coexist with the other two species 
and the cycle detaches from the $(x_2,y)$ plane. We can observe in Fig. \ref{ciclos} how these twist 
in the three-dimensional phase space,  displaying an oscillatory  coexistence of 
the three species. 
At $\mu_2\approx 0.56$ another Hopf bifurcation occurs, this time destroying the cycle, preserving the coexistence between the three species, as shown by the three rightmost points of Fig.~\ref{ciclos}B.
 
\begin{figure}[t]
\centering
\includegraphics[width=0.6\columnwidth]{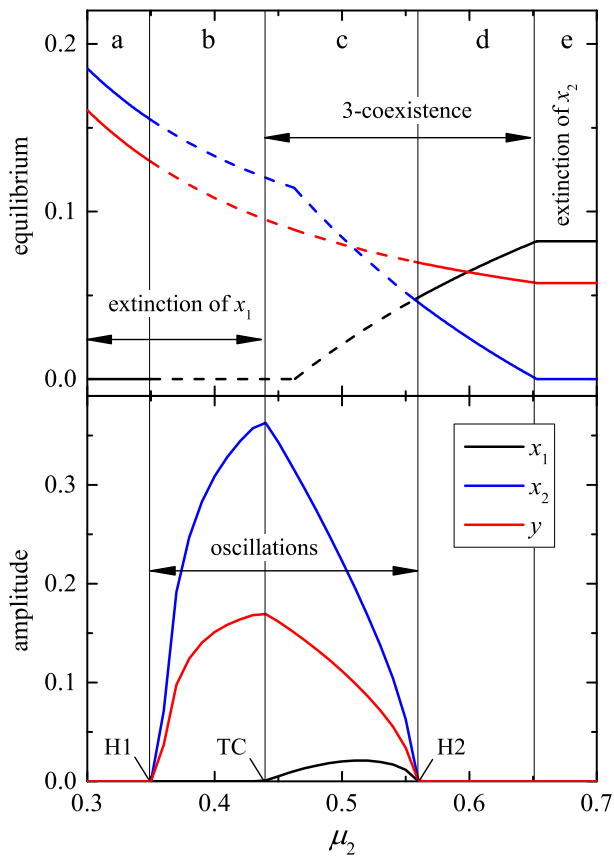}
\caption{Diagram of coexistence and extinction of the species described by Eqs.~(\ref{eq:mod}), as 
a function of the parameter $\mu_2$. Vertical lines separate the five different regimes observed, corresponding to the named panels of Fig.~\ref{evol}. 
Upper panel: equilibrium values (dashed lines show unstable equilibria). Lower panel: amplitude of 
the. Also shown are the two Hopf bifurcations, H1 and H2, and the transcritical bifurcation between two- and three-species cycles. All the remaining parameters are equal to those of Fig.~\ref{evol}.}
\label{diagram}
\end{figure}

A bifurcation diagram of the phenomenon, using $\mu_2$ as a control parameter, is shown in Fig.~\ref{diagram}, where the five regimes of Fig.~\ref{evol} are indicated by the same letters, in vertical stripes in both panels. The upper panel  displays the equilibria of the solutions. Dashed lines indicate linearly unstable equilibria, and in such circumstances sustained oscillations occur. The amplitude of these oscillations is shown in the bottom panel of Fig.~\ref{diagram}. 

We can observe more clearly that there is a region where species $x_2$ and the predator coexist (that is, with 
extinction of the dominante herbivore), corresponding to values of $\mu_2\lesssim 0.46$. This regime contains the Hopf bifurcation H1, with two-species oscillations for $\mu_2\gtrsim 0.36$. When the predation pressure 
on the inferior herbivore is increased above the transcritical bifurcation TC we observe coexistence of the three 
species, both in an oscillating regime and in a stationary equilibrium, achieved after the Hopf bifurcation H2 is crossed. Finally, if this predation is too high, it is $x_2$ the extinct species, allowing for the survival of the dominant species $x_1$.

\begin{figure}[ht]
\centering
\includegraphics[width=0.6\columnwidth]{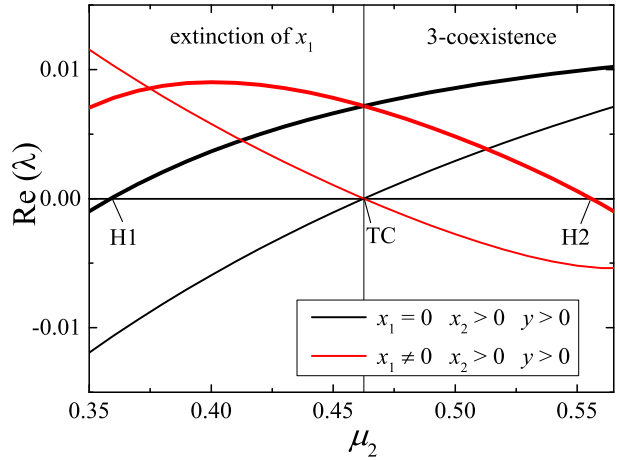}
\caption{Real part of the eigenvalues corresponding to the linear stability analysis of the equilibria of Eqs.~(\ref{eq:mod}) within the range of $\mu_2$ where oscillations are observed.
Thick lines correspond to  double eigenvalues. All the remaining parameters are equal to those 
of Fig.~\ref{evol}.}
\label{eigen}
\end{figure}

Complementing the bifurcation analysis, we show in Fig.~\ref{eigen} the real part of the 
eigenvalues of the linearized system at the unstable equilibria in the region of cycles, around the 
transcritical bifurcation TC. Thicker lines (of both colors) correspond to the pair of 
complex-conjugate eigenvalues of each cycle. Black lines correspond to the two-species oscillation, 
which is stable for $\mu_2\lesssim 0.462$. The eigenvalue with negative real part corresponds to the 
stable manifold of the cycle, which is normal to the plane $(x_2,y)$. At the transcritical 
bifurcation point TC this eigenvalue exchanges stability with the corresponding one of the other 
cycle (thin red line), the center manifold abandons the plane $x_1=0$ and three-species coexistence 
ensues.

\section{Final remarks and conclusions}
\label{conclusion}
 We have presented here the main results obtained with a simple three-species model, composed of a 
predator and two herbivores in asymmetric competition, where the predation pressure saturates if the 
density of preys is high enough. As shown, the model predicts the existence of different regimes as 
the values of the parameters change. These regimes consist of the survival of a single species (any 
of them), the coexistence of two species and the extinction of the third one (the three combinations 
are possible) and also the coexistence of the three species. But the most interesting aspect of the 
solutions of this model is that in some cases the populations oscillate in time. Under some 
conditions these oscillations are transient phenomena that decay to a stable equilibrium. Yet in 
other situations the oscillations are maintained indefinitely. In fact, we have found regions of 
coexistence of the three species with persistent oscillations of constant amplitude. 

It was shown that, while in the original model without saturation in the predation the asymptotic 
solutions converge to stable nodes, the steady state solutions of the model whith saciation shows 
both stable nodes and cycles. Our results indicate that, for low predation pressures on the inferior 
herbivore, the superior one extinguishes and non-oscillatory solutions appear for the remaining 
species, as indicated by the nodes observed in Fig.~\ref{ciclos}. At higher predation pressure a 
Hopf bifurcation and cycles develop, but still the superior herbivore cannot survive. After that, 
for an even higher value of $\mu_2$, 
a transcritical bifurcation of cycles occurs to a state of three-species coexistence. 
Bear in mind that the persistence of the inferior competitor requires that they have some advantage over the dominant one (in this case, a greater colonization rate). In such a context, the superior competitor is the most fragile of both with respect to predation (or to habitat destruction, as shown for example in \cite{11}). For this reason an increase of the predation on $x_2$ releases competitive pressure, allowing $x_1$ to survive. Finally, at an even higher predation pressure, another Hopf bifurcation occurs which destroys the cycle. The coexistence of the three species is preserved until the pressure $\mu_2$ is high enough to extinguish the inferior herbivore $x_2$.

Transcritical bifurcation of cycles in the framework of population models has been found in several 
systems described by equations that include saturation \cite{12,13,14,15,16,17}. Three-species food 
chain models were extensively studied through bifurcation analysis \cite{12,13,14}. A rich set of 
dynamical behaviors was found, including multiple domains of attraction, quasiperiodicity, and 
chaos. In Ref.~\cite{15} the dynamics of a two-patches predator-prey system is analyzed, showing 
that synchronous and asynchronous dynamics arises as a function of the migration rates. In a 
previous work, the same author analyzes the influence of dispersal in a metapopulation model 
composed of three species \cite{16}. Our contribution, through the model presented here, naturally 
extends those results in two directions. First, our model has three species in two trophic levels, 
two of them in asymmetric competition and subject to predation. Second, spatial extension and 
heterogeneity has been taken into account implicitly  as 
mean field metapopulations in the framework of Levins' model \cite{5}.  

Of course, we have not exhausted here all the possibilities of the model defined by Eqs.~(\ref{eq:mod}), but it is an example of the most interesting results that we have found. 
One can also imagine that the cyclic solutions arise from the interplay of activation and repression 
interactions, as in metabolic systems~\cite{monod1961}. The same pattern could be applied to 
regulations in community ecology if we replace the satiation inhibitor by the addition of a second 
predator, superior competitor with respect to the other predator, inhibiting its actions. This is a 
well documented pattern in several ecosystems~\cite{terborgh2010}. We believe 
that these behaviors are very general and will provide a thorough analysis elsewhere. These 
dynamical regimes considerably enrich the predictive properties of the model. In particular, we believe that the prediction of cyclic behavior for a range of realistic predator-prey models should motorize the search for their evidence in populations of current and extinct species. 


\section*{Acknowledgments}
The authors gratefully acknowledge grants from CONICET (PIP 2011-310), ANPCyT (PICT-2014-1558) and UNCUYO (06/506).

\end{document}